\documentclass[reprint,aps,pre,groupedaddress,showpacs]{revtex4-1}
\DeclareMathAlphabet{\mathpzc}{OT1}{pzc}{m}{it}
\usepackage{bm}
\usepackage{graphicx}
\usepackage{color}
\usepackage{amsmath}
\usepackage{natbib}
\usepackage{url}
\usepackage{textcase}
\usepackage{amsthm}
\usepackage{mathcomp}
\usepackage{amsfonts}
\usepackage{amssymb}
\usepackage{mathtools}

\begin{document}

\title{Integrable nonlinear parity-time symmetric optical oscillator}

\author{Absar U. Hassan}
\email{absar.hassan@knights.ucf.edu}
\author{Hossein Hodaei}
\author{Mohammad-Ali Miri}
\author{Mercedeh Khajavikhan}
\author{Demetrios N. Christodoulides}

\affiliation{CREOL/College of Optics and Photonics, University of Central Florida, Orlando, Florida 32816, USA}

\date{\today}

\begin{abstract}
The nonlinear dynamics of a balanced parity-time symmetric optical microring arrangement are analytically investigated. By considering gain and loss saturation effects, the pertinent conservation laws are explicitly obtained in the Stokes domain-thus establishing integrability. Our analysis indicates the existence of two regimes of oscillatory dynamics and frequency locking, both of which are analogous to those expected in linear parity-time symmetric systems. Unlike other saturable parity time symmetric systems considered before, the model studied in this work first operates in the symmetric regime and then enters the broken parity-time phase.
\end{abstract}


\maketitle

\section{\label{Sec1}Introduction}
The concept of parity-time ($\mathcal{PT}$) symmetry emerged within the framework of quantum field theories where it was found that Hamiltonians respecting this attribute could possess a real eigenvalue spectrum, despite being non-Hermitian~\cite{Bender1998}. A direct outcome of this possibility is the emergence of states that neither decay nor grow even in the presence of dissipation or gain~\cite{Znojil,*Ahmed,Makris,*Ramy}. In recent years, $\mathcal{PT}$-symmetric notions have attracted considerable attention and naturally led to research activity in many and diverse areas of physics that is still ongoing~\cite{Ruter,TKottosElectronicPT,AluAcoustics,*RamezaniPRX,EMGrafePTBoseHubbard,Yang_PhononPT}. Along these lines, optics provided a fertile ground where a series of intriguing phenomena related to $\mathcal{PT}$-symmetry can be directly observed by exploiting the mathematical isomorphism between the optical wave equation and the Schr\"{o}dinger equation ~\cite{LonghiPRL,Klaiman,Guo,Makris2,Stone,Sukov}. In the physical domain, this prospect was aided by the fact that amplification and attenuation of light can be effectively controlled in photonic structures.

For an optical potential to be $\mathcal{PT}$-symmetric, the complex refractive index distribution must obey the relationship $n(\bm{r})=n^*(-\bm{r})$ where $\bm{r}$ represents the position vector. This necessary (albeit not sufficient) condition implies that the real part of the index profile must be an even function in space while its imaginary counterpart, that is responsible for amplification/attenuation, should be odd~\cite{Makris}. These conditions demand that a $\mathcal{PT}$-symmetric structure must involve identical elements, e.g. two coupled cavities or waveguide elements, where gain and loss are anti-symmetrically distributed~\cite{Hodaei,*FengPTlaser,*Hodaei_OL,Feng3,MiriSoliton,MXiaoPTcoupled,*BPengPTcoupled,YangLossInduced,SRotterReversing}.

In general, optical configurations respecting $\mathcal{PT}$-symmetry exhibit two distinct phases. In the first one, the eigenvalue spectrum is purely real and thus no net amplification or decay of the field is expected to occur (exact $\mathcal{PT}$ phase). Instead, in the second one, some of the modes start to experience net growth or decay (in space or time) by entering the symmetry broken phase. The transition between these two regimes crucially depends upon the degree of non-Hermiticity (gain-loss contrast) and the coupling between adjacent sites~\cite{Ruter}. In addition, it is marked by the presence of an exceptional point where some of the eigenvalues and their respective eigenvectors tend to converge~\cite{Kato,Berry,WDHeiss,Soljacic}. At this point it is important to note that these results are direct byproducts of linear theories. Yet, in many optical realizations, nonlinearity is not only unavoidable but also often prevalent. This is particularly true in semiconductor-based systems where saturation effects strongly influence both gain and loss, to the point that a reversal in $\mathcal{PT}$-symmetry breaking can occur~\cite{AbsarPRA}. Clearly, of importance will be to understand at a fundamental level, the role such nonlinear processes play in the dynamics of $\mathcal{PT}$-symmetric arrangements~\cite{SegevNonlinearPT,Barashenkov1,*Barashenkov2, EMGrafeStationaryPT}.
\section{\label{Sec2}Dynamical model of the $\mathcal{PT}$-symmetric oscillator}
In many optical settings, nonlinearity typically manifests itself at high intensities by influencing the real as well as the imaginary part of the refractive index. In general, the imaginary component of the refractive index is nonlinearly modified through the presence of saturation effects in the effective gain or loss. In addition, the real part of the index also varies with intensity depending on whether the nonlinearity is of the focusing or defocusing type, as dictated by pumping conditions~\cite{Agrawal}. In semiconductor systems, gain saturation is responsible for clamping the light intensity within a resonator as well as the output power.

Here, we study the case where light density within a semiconductor structure remains below its saturation limit. This can be achieved by restricting the small-signal gain to relatively low values above the system loss. Moreover, in a travelling waveguide amplifier arrangement, the length of the device provides another degree of freedom in controlling the output optical intensity. Under these considerations, balancing field amplification and decay in an evanescently coupled structure composed of two identical elements, renders the system $\mathcal{PT}$-symmetric. In this respect, the optical/electrical pumping level in typical designs based on semiconductor quantum wells allows control over the values of both the gain and loss~\cite{Hodaei,AbsarPRA}, whereas the spatial separation between the components of the dimer determines the respective coupling strength.

In such a configuration the solution regimes are dictated by the gain (or loss) to coupling ratio which we here represent by $g\in\mathbb{R}^+$. By assuming that the linear losses due to scattering and absorption are small in comparison with the coupling strength, the field dynamics in the two components are found to obey the following dimensionless differential equations,
\begin{subequations}
 \label{full system1}
 \begin{align}
 \frac{d}{d\tau}u &= g\left(1-|u|^2\right)u + iv\label{coupled 1}
 \\
 \frac{d}{d\tau}v &= -g\left(1-|v|^2\right)v + iu\label{coupled 2}
 \end{align}
\end{subequations}

In the equations above, $u$ represents the field amplitude in the amplifying element while $v$ that in the lossy counterpart. Both $u$ and $v$ have been normalized with respect to a common gain-loss saturation value. The independent variable $\tau$ represents a spatial propagation coordinate (in the case of waveguide geometries) or time (in cavities), and is also scaled with respect to the coupling coefficient, $\kappa$. In a temporal representation involving a coupled micro-ring configuration, $\kappa$ is of the order $10^{11} \text{s}^{-1}$. Gain and loss in the presence and absence of pump light respectively, are also of the same order~\cite{AbsarPRA}. In what follows we determine the critical points of this nonlinear system and through the use of Stokes parameters, identify conservation laws and regimes of oscillatory and stationary responses.
\section{\label{Sec2}Critical Points and their stability}
Before we establish the integrability of Eqs.~(\ref{full system1}), it may be beneficial to first study the critical points involved and their associated stability properties. It is important to note that if $\left(u_0, v_0\right)$ represents a critical point, then so does $\left(u_0,v_0\right)e^{i\phi_0}$, where the phase $\phi_0$ is arbitrary. This leads to the conclusion that only the relative phase between the two complex quantities $\left(u_0, v_0\right)$ should be considered in the analysis. If we take for convenience $u_0$ to be real, it then follows from Eq.~(\ref{coupled 1}) that $v_0=i\rho u_0$ where $\rho\in\mathbb{R}$. In this case, under steady-state conditions, one finds that,
\begin{subequations}
 \label{RHOandU0}
 \begin{eqnarray}
 \left(\rho^2-1\right)\left(\rho^2-g\rho+1\right) = 0\label{RHO}
 \\
 u_0^2 = 1 - \frac{\rho}{g}\label{U0}
 \end{eqnarray}
\end{subequations}
The algebraic roots of Eq.~(\ref{RHO}), signifying the critical points, are given by $\rho=\pm1,(g\pm\sqrt{g^2-4})/2$. As we will show, among these four possible values of the modal ratio $\rho$, only one of them happens to be stable. To this end, linear stability analysis is carried out assuming small perturbations, i.e. $(u,v)\rightarrow(u_0+\varepsilon_1,i\rho u_0+\varepsilon_2)$ where in general $\varepsilon_{1,2}$ are complex. Upon substitution in Eqs.~(\ref{full system1}), we obtain the following differential equations concerning these perturbations,
\begin{subequations}
 \label{PERTRBS}
 \begin{eqnarray}
 \dot{\varepsilon_1}+g u_0^2(\varepsilon_1+\varepsilon_1^*)-g(1-u_0^2)\varepsilon_1-i\varepsilon_2=0\label{pertrb1}
 \\
 \dot{\varepsilon_2}+g \rho^2 u_0^2(\varepsilon_2^*-\varepsilon_2)+g(1-\rho^2 u_0^2)\varepsilon_2-i\varepsilon_1=0\label{pertrb2}
 \end{eqnarray}
\end{subequations}
These equations can be further simplified by using Eqs.~(\ref{RHOandU0}). Breaking down $\varepsilon_{1,2}$ in terms of their real $(\varepsilon_{1R,2R})$ and imaginary $(\varepsilon_{1I,2I})$ parts, one obtains an eigenvalue equation, $MX=\lambda X$ by assuming a temporal dependence of the form $e^{\lambda \tau}$. Here $X$ represents the eigenvector, $X=(\varepsilon_{1R_0},\varepsilon_{1I_0},\varepsilon_{2R_0},\varepsilon_{2I_0})^T$ of the matrix $M$,
\[ M = \left( \begin{array}{cccc}
3\rho-2g   & 0      & 0         & -1 \\
0           & \rho   & 1         &  0 \\
0           & -1     & -1/\rho   & 0 \\
1           & 0      & 0         & -(3/\rho-2g)\end{array} \right).\]
The characteristic equation for this system is given by the following expression,
\begin{multline}
 \label{characteristic}
 \lambda \left\{\lambda^2-3(\rho-\frac{1}{\rho})\lambda+1-(2g-3\rho)(2g-\frac{3}{\rho})\right\}\\
 \left\{\lambda-\left(\rho-\frac{1}{\rho}\right)\right\} = 0
\end{multline}
We next separately analyze the stability properties of the four stationary points $\rho$. In this case we find:

(i) $\rho=-1$.
\begin{align*}
 \lambda_{1,2} &= 0
 \\
 \lambda_{3,4} &= \pm2\sqrt{(g+1)(g+2)}
\end{align*}
Since $g>0$, there always exists a positive real $\lambda$ which renders this point unstable.

(ii) $\rho=+1$.
\begin{align*}
 \lambda_{1,2} &= 0
 \\
 \lambda_{3,4} &= \pm2\sqrt{(g-1)(g-2)}
\end{align*}
For $g<1$ or $g>2$, it is clear that one eigenvalue is a positive real number so that this point becomes unstable. Moreover, the double eigenvalue $(\lambda_{1,2}=0)$ is not semi\textendash simple~\cite{Seyranian_multiparameter}(also true for $1<g<2$) and leads to terms proportional to $\tau$ in the general solution, thus introducing instability.

(iii) $\rho=(g+\sqrt{g^2-4})/2$.
\begin{align*}
 \lambda_1 = 0,  &\quad \lambda_2 = \sqrt{g^2-4}
 \\
 \lambda_{3,4} &= 0.5(3\pm1)\sqrt{g^2-4}
\end{align*}
If $g>2$, then all eigenvalues are positive and hence this stationary point is unstable. On the other hand, for $g<2$, $\rho$ is complex, hence violating Eq.~(\ref{U0}) for $u_0^2$.

(iv) $\rho=(g-\sqrt{g^2-4})/2$.
\begin{align*}
 \lambda_1 = 0,  &\quad \lambda_2 = -\sqrt{g^2-4}
 \\
 \lambda_{3,4} &= -0.5(3\pm1)\sqrt{g^2-4}
\end{align*}
Stability is here ensured for $g>2$ since all the eigenvalues are negative ($\lambda\leq0$). On the other hand if $g<2$, this point does not exist for the same reason as mentioned in the previous case. Note that the critical point corresponding to the value of $g=2$ makes the cases (ii)-(iv) equivalent and is found to be stable.

\begin{figure}
  \includegraphics{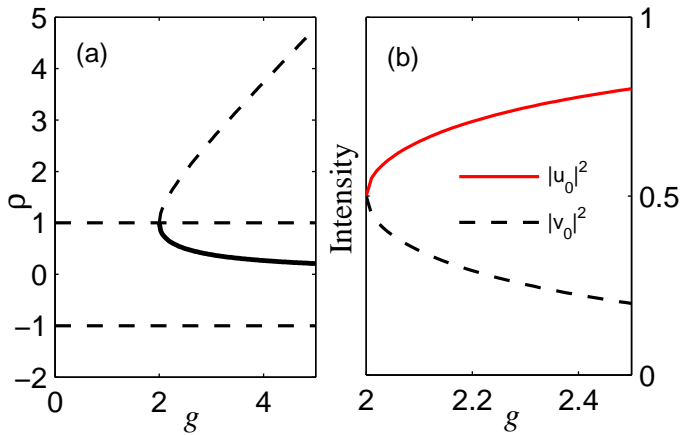}
  \caption{\label{Fig1}The various branches of the ratio $\rho$ associated with the critical points as a function of $g$ are displayed in (a) where the solid line indicates stable behavior while the dashed unstable. (b) Intensities in the two optical elements corresponding to the stable critical point are plotted as the value of $g$ increases.}
\end{figure}

A bifurcation diagram describing the behavior of the critical points as a function of the gain-loss constant ($g$) is shown in Fig.~\ref{Fig1}(a), where the stable branch of $\rho$ is depicted as a solid line. As the value of gain increases beyond $g=2$, the ratio between the fields starts decreasing, starting from $\rho=1$ and asymptotically reaching $\rho=0$. This behavior is reminiscent of linear $\mathcal{PT}$-symmetric systems where in the broken symmetry domain (after a bifurcation in the eigenvalues beyond an exceptional point) the field strengths in the gain and loss components becomes unequal. This is shown in Fig.~\ref{Fig1}(b) for $g>2$. However, in contrast with a linear $\mathcal{PT}$-symmetric dimer where an exponential increase in intensities is expected with time, the saturation in Eq.~\ref{full system1} will enforce a bounded steady-state for $g>2$. Furthermore, once this $\mathcal{PT}$-symmetry is broken, light tends to predominantly reside in the cavity that offers amplification, as the gain-loss contrast is increased.
Moreover, the stability of the trivial critical point at the origin ($u_0=v_0=0$) needs also to be considered. Here, the differential equations for the perturbations assume the following form,
\begin{subequations}
 \label{PERTRBS-origin}
 \begin{eqnarray}
 \dot{\varepsilon_1}-g\varepsilon_1-i\varepsilon_2=0\label{pertrb1-origin}
 \\
 \dot{\varepsilon_2}+g\varepsilon_2-i\varepsilon_1=0\label{pertrb2-origin}
 \end{eqnarray}
\end{subequations}
Again using the representation, $(\varepsilon_{1},\varepsilon_{2})=(\varepsilon_{01},\varepsilon_{02})e^{\lambda t}$, the eigenvalues of this system are found to be $\lambda_{1,2}=\pm\sqrt{g^2-1}$. In the range $g<1$, these values are purely imaginary and conjugate to eachother, thus implying an unstable saddle point. On the other hand, for $g>1$, there exists a positive real $\lambda$ indicating an unstable exponential growth.
\begin{figure}
  \includegraphics{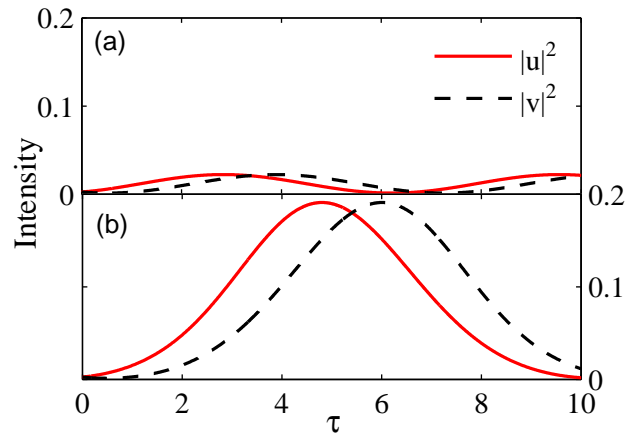}
  \caption{\label{Fig2}The effect of the linear $\mathcal{PT}$-symmetry breaking around $g=1$, is depicted. (a) For $g=0.9$, sinusoidal oscillations occur while (b) for $g=1.1$, an exponential growth takes place until saturation comes into play.}
\end{figure}
However, it is instructive to notice that when fields in both cavities start from noise where $|u(0)|,|v(0)|\sim0$, the dynamics reduce to that of a linear $\mathcal{PT}$-symmetric coupler, governed by Eqs.~(\ref{PERTRBS-origin}), with $\varepsilon_{1}$ and $\varepsilon_{2}$ being replaced with $u$ and $v$ respectively. In this linear scenario, it is well known that the $\mathcal{PT}$-symmetric phase transition occurs at the point where the gain-loss to coupling ratio is unity. The role of this spontaneous symmetry breaking point at $g=1$ is apparent in Fig.~\ref{Fig2} where the initial values were chosen to be small $|u(0)|,|v(0)|=10^{-2}$. Below this breaking point, the intensities evolve sinusoidally\textemdash characteristic of unbroken symmetry eigenmodes~\cite{Ruter}; For $g>1$, the linear symmetry breaks and an initial exponential growth occurs up to the point where the intensities get larger and saturation starts to limit this growth.
\section{\label{Sec3}Stokes Parameters}
\begin{figure*}
  \includegraphics{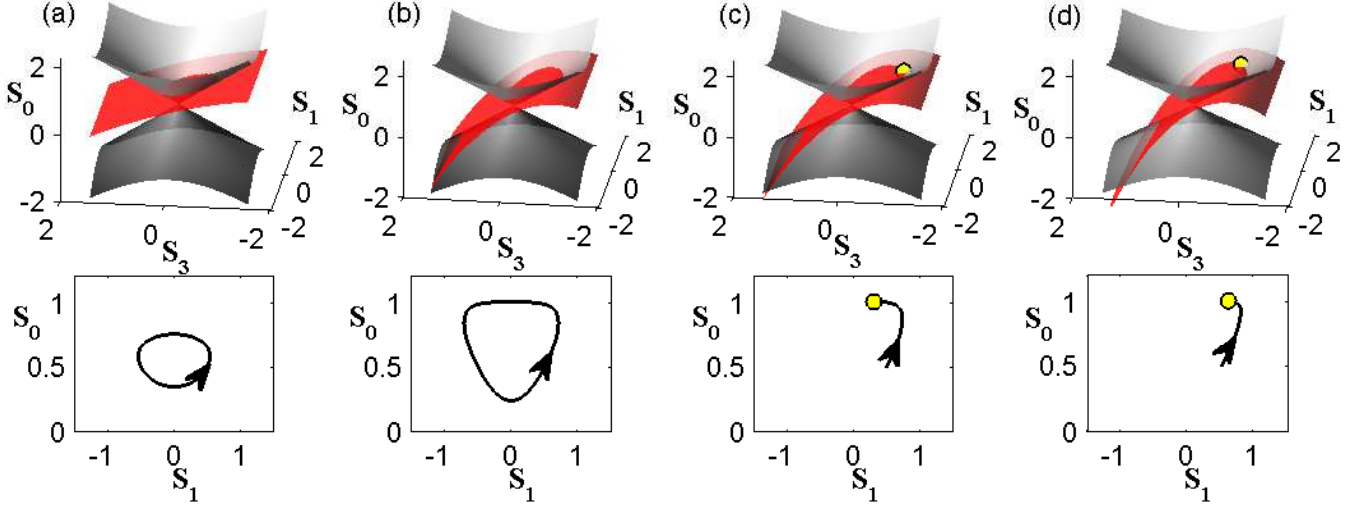}
  \caption{\label{Fig3}Intersections between two surfaces in the $(S_1,S_3 S_0)$-space are plotted that describe the solution trajectories. These are shown for two values of $g$ both below, (a) $g=0.8$ and (b) $g=1.9$, and above, (c) $g=2.1$ and (d) $g=2.5$, the nonlinear phase transition point at $g=2$. Corresponding plots in the lower panel depict the intersections in the $(S_1,S_0)$ plane. The stable critical point appears for $g>2$ and is shown as a yellow dot. In all cases, initial values of the fields are $u(0)=0.7(1+0.1i)$ and $v(0)=0$.}
\end{figure*}
In this section we analyze the properties and behavior of this non-Hermitian nonlinear dynamical system Eq.~\ref{full system1} using Stokes parameters. To do so, we first obtain the conservation laws that are needed to establish integrability. The Stokes parameters are defined as follows,
\begin{subequations}
 \label{STOKES}
 \begin{align}
 S_0&=|u|^2+|v|^2\label{Stk0}
 \\
 S_1&=|u|^2-|v|^2\label{Stk1}
 \\
 S_2&=u^*v+uv^*\label{Stk2}
 \\
 S_3&=i(u^*v-uv^*)\label{Stk3}
 \end{align}
\end{subequations}
These four real quantities listed here are all real and are interrelated by the expression,
\begin{equation}
S_0^2=S_1^2+S_2^2+S_3^2\label{Power}
\end{equation}
The dynamical equations for each of these four parameters can be directly obtained using Eq.~(\ref{full system1}), i.e.,
\begin{subequations}
 \label{STOKES-derv}
 \begin{align}
 \dot{S_0}&=-2gS_0S_1+2gS_1\label{Stk0_derv}
 \\
 \dot{S_1}&=-g(S_0^2+S_1^2)+2gS_0+2S_3\label{Stk1_derv}
 \\
 \dot{S_2}&=-gS_1S_2\label{Stk2_derv}
 \\
 \dot{S_3}&=-(2+gS_3)S_1\label{Stk3_derv}
 \end{align}
\end{subequations}
From Eqs.~(\ref{Stk0_derv}), Eq.~(\ref{Stk2_derv}) and Eq.~(\ref{Stk3_derv}), one can establish that, $-S_1=\dot{S_2}/(gS_2)=\dot{S_3}/(2+gS_3)=\dot{S_0}/(2gS_0-2g)$, which immediately leads to the following two conservation laws,
\begin{align}
 A&=\frac{S_2}{2+gS_3}\label{CoM-A}
 \\
 B&=\frac{S_2^2}{S_0-1}\label{CoM-B}
 \end{align}
Clearly, the existence of these two constants of motion implies integrability. These two constants are determined by the initial values of the Stokes parameters and the gain-loss contrast. To find the evolution trajectory of $u(\tau)$ and $v(\tau)$, it suffices to know the dynamics of only one Stokes parameter. In this case, by first expressing $S_3$ and $S_0$ in terms of $S_2$ [using Eqs.~(\ref{CoM-A}) and ~(\ref{CoM-B})], in Eq.~(\ref{Power}) and finally using Eq.~(\ref{Stk2_derv}) that relates $S_1$ to $S_2$ and $\dot{S_2}$, we obtain a differential equation solely involving $S_2$,
\begin{equation}
\label{SingleSDynamic}
(\dot{S_2})^2=g^2S_2^2\left\{\left(1+\frac{S_2^2}{B}\right)^2-S_2^2-\frac{1}{g^2}\left(\frac{S_2}{A}-2\right)^2\right\}
\end{equation}
In principle, Eq.~(\ref{SingleSDynamic}) can be solved by quadrature. Hence from $S_2(\tau)$, $S_3(\tau)$ and $S_0(\tau)$ can then be recovered through the conservation laws and finally $S_1(\tau)$ can be found using Eq.~(\ref{Stk2_derv}) or Eq.~(\ref{Power}). This enables the dynamics of all four Stokes parameters to be determined. From here one can obtain the original field amplitudes and phases via Eqs.~(\ref{STOKES}), e.g. $|u(\tau)|^2=(S_0(\tau)+S_1(\tau))/2$.

The trajectories followed by the solutions can be conveniently described through plots in the Stokes space of $(S_1,S_3,S_0)$. These are governed by intersections between a hyperboloid and a parabola, as dictated by Eqs.~(\ref{Power}),~(\ref{CoM-A}) and (\ref{CoM-B}),
\begin{align}
 \left(B-\frac{B^2}{4}\right)&=S_1^2+S_3^2-\left(S_0-\frac{B}{2}\right)^2\label{Hyperbola}
 \\
 S_0&=1+\frac{A^2}{B}(2+gS_3)^2\label{Parabola}
\end{align}
Following this approach, it is possible to determine the domains pertaining to instability, i.e. the conditions leading to open-ended intersections or trajectories. But when this system is initiated within the linear regime ($|u(0)|^2,|v(0)|^2\ll 1$) no such domains of instability were identified as $g$ was varied.
\begin{figure}
  \includegraphics{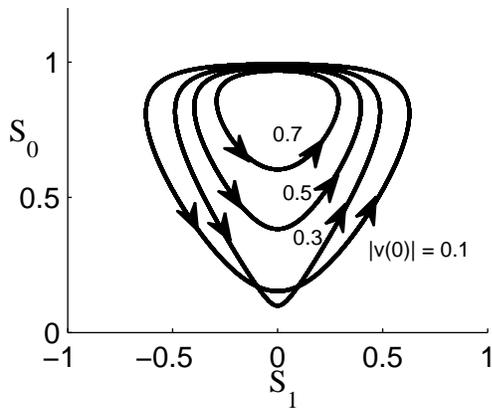}
  \caption{\label{Fig4}Different solution trajectories in the $(S_1,S_0)$-space are shown as the initial conditions are changed. For these plots, $u(0)$ is fixed at $u(0)=0.7$, while $v(0)$ is varied in the imaginary space from $v(0)=0.1i$ to $v(0)=0.7i$. The gain-loss value used is $g=1.8$.  Arrows indicate the evolution over time.}
\end{figure}
To explore the behavior of the system we chose to map the Stokes dynamics on the hyperboloid of Eq.~(\ref{Hyperbola}) since it is independent of $g$. The two surfaces are plotted for four different values of $g$ in Fig.~\ref{Fig3}. For $g<2$, the Stokes parameters follow periodic trajectories. On the other hand, when $g>2$, the intersection of the parabola and hyperboloid passes through the critical point which is stable under these conditions [(iv)), Sec.~\ref{Sec2}]. Here, instead of a periodic evolution, the field values attain a steady state of unequal values analogous to that occurring in a $\mathcal{PT}$-symmetry broken scenario. This case is shown in part (c) and (d) of Fig.~\ref{Fig3}.

Note that the solution profiles depicted in Fig.~\ref{Fig3}(a) and ~\ref{Fig3}(b) indicate the presence of oscillations akin to stable limit cycles. On the other hand, by changing the initial conditions (keeping $g$ fixed), these orbits become modified. This implies that these are not exactly limit cycles but instead neutrally stable cycles. To demonstrate this, in Fig.~\ref{Fig4} we set the field in the cavity with gain to be $u(0)=0.7$ and we then increase the initial value of the field in the cavity with loss. In this case, the cycles in the $(S_1,S_0)$ space are found to change accordingly.

Considering the results presented, one can infer the existence of two distinct responses associated with Eq.~(\ref{full system1}). The first corresponds to solutions expected in a system like the well-known Van der Pol oscillator ~\cite{VanDerPol}. This domain is defined by $g<2$, and here the intensities in both cavities behave in a very similar manner (reflected versions of each other) having the same period and lying within an identical bounded interval. Whereas in the second regime, the fields are pulled into the stable critical point given in part (iv) Sec.~\ref{Sec2}. The former relates to the $\mathcal{PT}$-symmetric phase since $|u|^2$ and $|v|^2$ oscillate symmetrically over time, while the latter is analogous to the symmetry broken phase where the two intensities are unequal.
\begin{figure}
  \includegraphics{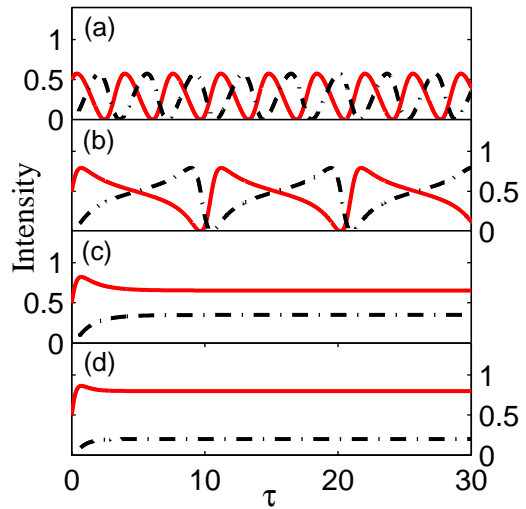}
  \caption{\label{Fig5}The behavior of the intensities over time in the two cavities is shown. The four graphs correspond to the four values of $g$ used in Fig.~\ref{Fig3}. Parts (a) $g=0.8$, and (b) $g=1.9$, depict a Van der Pol-like oscillatory regime and (c) $g=2.1$, and (d) $g=2.5$, the $\mathcal{PT}$-broken phase. Solid (red) lines corresponds to intensity in the component with gain and dashed (black) to that in the component with loss.}
\end{figure}
Numerical results from a Runge-Kutta simulation for these two phases (corresponding to values of $g$ in Fig.~\ref{Fig3}) are depicted in Fig.~\ref{Fig5} where ~\ref{Fig5}(a) and ~\ref{Fig5}(b) show intensities in the symmetric domain and ~\ref{Fig5}(c) and ~\ref{Fig5}(d) display the broken phase. In this latter scenario, we also found that the fields in both components of the dimer are locked at the common resonant frequency (or propagation constant) of the cavities (or waveguides) \textemdash a feature of spontaneously broken $\mathcal{PT}$-symmetry. Another characteristic of this $\mathcal{PT}$-phase can be deduced from the fact that as $g$ increases, the ratio $|v_0/u_0 |^2$ becomes gradually smaller. In addition, once the system starts to oscillate within the symmetric regime, the transition between the two domains occurs at the nonlinear boundary $g=2$ as the gain-loss value is increased. This is in contrast to a linear $\mathcal{PT}$-symmetric coupler where the transition occurs instead at $g=1$. Although nonlinear saturation effects tend to modify the location of this transition in the parameter space, the order in which it takes place is not affected - unlike in other nonlinear $\mathcal{PT}$-symmetric settings~\cite{AbsarPRA}.

Finally, an interesting feature associated with this oscillator is the fact that within the exact $\mathcal{PT}$-symmetry domain, as the system gets close to the nonlinear phase transition point, the period of oscillations tends to approach infinity. Now consider operation close to $g=2$, for instance in a coupled micro-ring resonator configuration when the gain-loss contrast between the rings is twice the coupling between them. This could lead to periodic flashes of light observable at much longer time scales compared to coupling times which are typically on the order of picoseconds.
\section{\label{Sec4}Conclusions}
In conclusion, we have investigated the behavior of a fully integrable non-Hermitian oscillator with a balanced gain-loss distribution. Our analysis indicates the existence of two regimes of oscillatory dynamics and frequency locking, both of which are analogous to those expected in linear $\mathcal{PT}$-symmetric systems. The oscillator was found to first operate in the symmetric regime before entering the broken $\mathcal{PT}$-phase at higher gain-loss values. Our study can shed light on the interplay of $\mathcal{PT}$-symmetry and nonlinearity.
\begin{acknowledgements}
The authors gratefully acknowledge the financial support from NSF CAREER Award (ECCS-1454531), NSF (grant ECCS-1128520 and DMR-1420620), AFOSR (grant FA9550-14-1-0037) and ARO (grant W911NF-14-1-0543 and W911NF-16-1-0013).
\end{acknowledgements}

\bibliography{REFS}

\end{document}